\documentclass[conference]{IEEEtran}
\usepackage{keyval}
\usepackage{subcaption}
\IEEEoverridecommandlockouts
\usepackage{amssymb}
\usepackage{amsmath}
\usepackage{graphicx}
\usepackage{amsmath,amsthm}
\usepackage{float}
\usepackage{enumerate}
\usepackage{epstopdf}
\usepackage{dblfloatfix}
\usepackage{caption}
\usepackage{amsfonts}
\usepackage{fancyhdr}
\usepackage{bbm}
\usepackage{cases}
\usepackage{cite}
\usepackage{breqn}
\usepackage{multirow}
\usepackage{mathrsfs}
\usepackage{algpascal}
\usepackage{algc}
\usepackage{algorithmicx}
\usepackage[ruled]{algorithm}
\usepackage{algpseudocode}
\usepackage{graphics}
\usepackage{epsfig}
\usepackage{mathrsfs} 
\usepackage{dsfont}
\usepackage{tabularx}
\usepackage{color}
\usepackage{booktabs}
 \usepackage{hyperref}

\floatname{algorithm}{\small{Algorithm}}

\algdef{SE}[DOWHILE]{Do}{doWhile}{\algorithmicdo}[1]{\algorithmicwhile\ #1}%

\ifCLASSINFOpdf
 \else
\fi

\begin{document}
\theoremstyle{definition} 
\newtheorem{theorem}{Theorem}[section]
\newtheorem{definition}[theorem]{Definition}
\newtheorem{lemma}[theorem]{Lemma}
\newtheorem{example}[theorem]{Example}
\newtheorem{Proposition}[theorem]{Proposition}
\newtheorem{Corollary}[theorem]{Corollary}

\title{Rate-Splitting Multiple Access for RIS-Enabled Full-Duplex Communications with User Relaying}

 

\author{
 \IEEEauthorblockN{ 
        Yuwen Cao\IEEEauthorrefmark{1}, 
        Hui Wang\IEEEauthorrefmark{1},
        Tomoaki Ohtsuki\IEEEauthorrefmark{2}, 
        and Shahid Mumtaz\IEEEauthorrefmark{3}
    }
    \IEEEauthorblockA{\IEEEauthorrefmark{1}College of Information Science and Technology, Donghua University, Shanghai, China}
    \IEEEauthorblockA{\IEEEauthorrefmark{2}Department of Information and Computer Science, Keio University, Yokohama, Japan}
\IEEEauthorblockA{\IEEEauthorrefmark{3}Department of Engineering, Nottingham Trent University, UK}
}

{}

\maketitle

\begin{abstract}
 Rate splitting multiple access (RSMA) has emerged as a promising non-orthogonal transmission mechanism to improve spectral- and energy efficiency, and user fairness in multi-antenna broadband channels. To further enhance its performance, in this paper we investigate a novel cooperative RSMA (C-RSMA) transmission framework assisted by reconfigurable intelligent surface (RIS) and user relaying. More specifically, we first formulate the common- and private-rate maximization problem satisfying the transmit power constraint, by jointly optimizing the transmit beamformers at the BS and the reflective beamformer at the RIS. The above formulated problem is an intractable and NP-hard problem as the optimization variables are coupled and the strong residual self-interference (SI) is involved. To conquer the challenge mentioned above, we propose a low-complexity alternative optimization approach to jointly optimize the transmit and reflective beamforming. 
 Simulation results show that our approach 
 is capable of enabling superior sum-rate performance than the RIS-assisted space division multiple access (SDMA), half-duplex RSMA, and SDMA without using RIS schemes. 
\end{abstract}


\section{Introduction}
As a novel non-orthogonal transmission framework and interference management strategy for multi-user multiple-input multiple-output (MU-MIMO)  communications, rate splitting multiple access (RSMA) has emerged as one of the key enabling technologies for current and future mobile networks~\cite{7470942}. RSMA has shown its potential to improve data rate, energy and spectral efficiency,
and robustness against imperfect channel state information (CSI) \cite{Mao2018}. One notable feature of RSMA is that it splits user messages of each data stream into i) common and ii) private parts. Then, the common parts are combined to one common message which is encoded in a common data stream by using the codebook shared by all users, whereas the private parts are encoded in a private data stream by using the codebook known merely to the associated user. 

On the other hand, the reconfigurable intelligent surface (RIS) \cite{11145277,basar2019wireless,11711120,10509625,11457010}, as another enabling technique for 6G communication networks, has introduced to further improve the communication performance in terms of data rate, energy and spectral efficiency, etc. The principle of RIS is to support the transmission from a source to a destination by dynamically reconfiguring the reflection coefficients, i.e., reconfigure the RIS reflection coefficients to beamform its received signal towards the desired destination \cite{8888223}. In such a way, RIS can help the communication system adapt to the propagation environment dynamically. Motivated by the appealing performance benefits of RSMA and RIS techniques, several researchers recently have investigated the integration of RSMA and RIS attempting to enhance the performance in terms of the weighted ergodic sum-rate (WESR) \cite{9663192}, user fairness\cite{9473650}, signal-to-interference-plus-noise ratio (SINR) coverage rate \cite{10757887}, and outage probability \cite{9508885}, etc. 
Moreover, some works have studied the interplay of RSMA and non-orthogonal multiple access (NOMA) techniques in terms of the data rate, sum degree-of-freedom (DoF) and quality of service (QoS), etc.
However, most of the aforementioned works only consider the downlink RIS-enabled RSMA framework design. Hence, it is 
critical to explore the joint optimization for both uplink and downlink transmissions of full-duplex (FD) cooperative RSMA (C-RSMA) framework assisted by the RIS to further enhance the data rate with low complexity.
     
In this paper, we investigate the RIS-assisted C-RSMA transmission framework, in which the RIS dynamically reflects the encoded common stream and private stream from both the base station (BS) and the relaying user (i.e., the helper) simultaneously.
More specifically, we first formulate a common- and private-rate maximization problem 
satisfying the transmit power
constraint, the above formulated problem is an intractable and NP-hard problem as the optimization variables are coupled and the strong residual self-interference (SI) is involved. To address this challenge, we propose a low-complexity alternative optimization approach that jointly optimizes the transmit beamformers at the BS and the reflective beamformer at the RIS. Notably, the major technical contributions are summarized as follows:

\begin{itemize}
\item We investigate the interplay between the C-RSMA and RIS in terms of common- and private-rate performance under the transmit power constraint.
    \item Unlike orthogonal multiple access (OMA), which requires separate time slots and necessitates designing RIS beamforming for each slot, our C-RSMA framework enables the simultaneous reflection of both common and private streams from the BS and the relaying user through the RIS. This concurrent reflection allows for a single RIS beamforming design, thereby reducing system complexity and overhead.
    \item Simulation results demonstrate that our proposed C-RSMA transmission system yields common- and private-rate performance gain up to 31.6 \% than the state-of-the-art half-duplex (HD) RSMA transmission system when the transmit power at the BS is 30 dBm. In addition, we observe that with given information transmission rate, C-RSMA consumes lower transmit power than the space division multiple access (SDMA) and HD-RSMA schemes assisted by RIS.\footnote{{\it Notations}: Bold lowercase and uppercase letters are used to represent vectors and matrices, respectively. $\left( {\cdot} \right)^{T}$, $\left( {\cdot} \right)^{H}$, $\mathrm{Tr}\left( {\cdot} \right)$, and $\left( {\cdot} \right)^{\star}$ denote the transpose, complex conjugate, the trace operations of a matrix, and the complex conjugate of a complex
variable, respectively. 
$\operatorname{diag}(\cdot)$ represents a matrix of the diagonal elements, and $\otimes$ indicates the Hadamard product operation. 
$\mathbb{C}$ denotes the complex space. $\mathbb{E}\{\cdot\}$ and $\operatorname{vec}(\cdot)$ mean the expectation and vectorization of the argument, respectively. $\mathcal{CN}(\mu, \boldsymbol{\Sigma})$ stands for the multivariate complex Gaussian
distribution with mean $\mu$ and covariance matrix $\boldsymbol{\Sigma}$.}
\end{itemize}

\begin{figure}
\setlength{\abovecaptionskip}{-0.0cm}
\setlength{\belowcaptionskip}{-0.4cm}
  \begin{center}
    \includegraphics[width=0.4\textwidth]{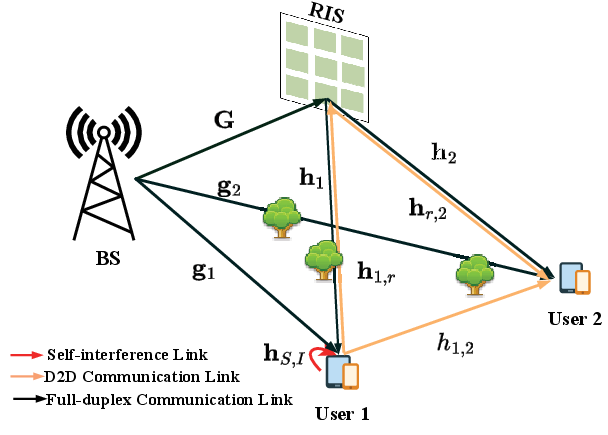}
    \end{center}
    \caption{An overview of the proposed C-RSMA transmission framework assisted by the single-connected RIS.} \label{fig:01}
\end{figure}

\section{System Model}
Consider a multi-user multiple-input single-output (MU-MISO) communication network in time division duplex (TDD) mode, comprising of one BS equipped with $M_t$ transmit antennas and two single-antenna users. In addition, our communication system involves the assistance of RIS with a group of $N$ passive reflecting elements indexed by $\mathcal{N} =\left\{1,2,\ldots,N\right\}$. Without loss of generality, we assume that the channel quality of user $1$ is better than that of user $2$, which is reflected in the model that user $1$ is closer to the BS whereas user $2$ is father away from the BS; 
An extension of the RIS-assisted C-RSMA transmission framework with multi-user multi-antenna scenario will be studied in our future research.     

As illustrated in Fig. \ref{fig:01}, in our C-RSMA transmission framework, user $1$ 
is selected to act as an FD relay to assist the BS in transmitting the common stream signal to user $2$, thereby improving the quality of the common stream for the user $2$. The BS simultaneously serves the two users with the assistance of a single-connected RIS. At the same time, user~1 is also transmitting communication signal to user 2 assisted by a single-connected RIS. 
Let $\mathbf{g}_k\in \mathbb{C}^{M_t\times 1}$, $\mathbf{G}\in \mathbb{C}^{N\times M_t}$, $\mathbf{h}_k\in \mathbb{C}^{N\times 1}$, $h_{\text{S,I}}$, and $h_{1,2}$ denote the channels between BS and users, the channel between BS and RIS, the channel between RIS and users, the residual SI channel, as well as the device-to-device (D2D) communication channel, respectively. 

The BS transmits signal to two users based on the principle of 1-layer RSMA. Meanwhile, these signal will also be transmitted to users through the reflection of RIS. 
We denote $\boldsymbol{s} = \left[ s_0, s_1, s_2 \right]^T\in \mathbb{C} ^{3\times 1}$ and $\boldsymbol{W} = \left[ \mathbf{w}_0, \mathbf{w}_1, \mathbf{w}_2 \right] \in \mathbb{C} ^{{M}_{{t}}\times 3}$ as the data stream vector and beamforming matrix for all streams, respectively. We assume each stream $s_l$, $l\in \left\{0,1,2 \right\}$, has zero mean and unit variance, i.e., $\mathbb{E} \left\{s_{l}^{*}s_{l}\right\} =1$. Therefore, the transmitted signal at the BS can be  expressed as
 $   \mathbf{x} =\sum\nolimits_{k=0}^2{\mathbf{w}_{l}{s}_{l}}$.
Besides, the transmit power constraint at BS is given by
$\mathrm{Tr}( \mathbf{WW}^{\mathrm{H}} ) \leq {P}_{t}$.
Moreover, $P_t$ is the maximum transmit power of the BS. The signal received at user 1 is not only transmitted directly through the BS, but also reflected by the RIS. Apart from that, the received signal at user 1 is embedded by undesired residual SI,
which in principle is a critical issue in FD operation.
Hence, the total received signal at user 1 can be expressed as 
\begin{equation}
\label{eq:03}
y_1=\left( \mathbf{g}_{1}^{H}+\mathbf{h}_{1}^{H}\boldsymbol{\Theta }\mathbf{G} \right) \mathbf{x}+\sqrt{P_{r}}h_{\text{S,I}}^{*}s_0+n_1.
\end{equation}
In (\ref{eq:03}), the RIS scatting matrix $\boldsymbol{\Theta }$ satisfies the constraint of
$\boldsymbol{\Theta }=\text{diag}\big( e^{j\theta _1},e^{j\theta _2},\ldots,e^{j\theta _N} \big)$,
where $\theta _n\in \left[ 0,2\pi \right) ,\ n\in \left\{ 1,2,\ldots ,N \right\}$,
and $n_1$ is the additive white Gaussian noise (AWGN) satisfying $n_1\thicksim \mathcal{C}\mathcal{N}\left( 0,\sigma _{1}^{2} \right) $. In this system, user $1$ transmits the decoded common stream $s_0$ to user $2$. Therefore, user $1$ receives an interference through the channel $h_{\text{S,I}}$. Besides, $P_r$ represents the maximum transmit power of the user 1 for D2D communication with the far away located user (i.e., the user 2) via both the reflective line-of-sight (LoS) link and the direct LoS link. 

At user 1, she first decodes the common stream $s_0$ by treating all private streams as interference. Thus, the SINR with self-interference of $s_0$ at user 1 can be expressed as 
\begin{equation}
\label{eq:005}
\gamma_{c,1}=\frac{\left| \left( \mathbf{g}_{1}^{H}+\mathbf{h}_{1}^{H}\Theta \mathbf{G} \right) \mathbf{w}_0 \right|^2}{\sum\nolimits_{i=1}^2{\left| \left( \mathbf{g}_{1}^{H}+\mathbf{h}_{1}^{H}\Theta \mathbf{G} \right) \mathbf{w}_i \right|}^2+P_{r}\left| h_{\text{S,I}} \right|^2+\sigma _{1}^{2}}.
\end{equation}
Therefore, the corresponding transmission rate of decoding the common stream at user 1 yields
$
r_{c,1}=\log _2\big( 1+\gamma_{c,1} \big)$.
After decoding the common stream $s_0$, user 1 removes the common stream $s_0$ by employing the successive interference cancellation (SIC) technique. Hence, the SINR at user $1$ decoding the intended private stream $s_1$ is equal to 
\begin{equation}
\gamma_{p,1}=\frac{\left| \left( \mathbf{g}_{1}^{H}+\mathbf{h}_{1}^{H}\boldsymbol{\Theta }\mathbf{G} \right) \mathbf{w}_1 \right|^2}{\left| \left( \mathbf{g}_{1}^{H}+\mathbf{h}_{1}^{H}\boldsymbol{\Theta }\mathbf{G} \right) \mathbf{w}_2 \right|^2+ P_{r}\left| h_{\text{S,I}} \right|^2+\sigma _{1}^{2}},
\end{equation}
and the corresponding rate of decoding the private stream is
$
r_{p,1}=\log _2\left( 1+\gamma_{p,1}\right)$.

Simultaneously, the signal received at user 2 comprises i) direct transmission from BS, ii) transmission reflected by the RIS from BS, iii) direct D2D transmission from user 1, and iv) transmission reflected by the RIS from user 1. Then, the total received signal at user 2 can be expressed as 
\begin{equation}
y_2=( \text{g}_{2}^{H}+\mathbf{h}_{2}^{H}\mathbf{\Theta G} ) \mathbf{x}+\sqrt{P_{r}}( h^{}_{1,2}+\mathbf{h}_{r,2}^{H}\mathbf{\Theta h}_{1,r} ) s_0 + n_2,
\end{equation}
where ${h}_{1,2}\in \mathbb{C}^{1\times 1}$, $\mathbf{h}_{r,2}\in \mathbb{C}^{{N}\times 1}$, and $\mathbf{h}_{1,{r}}\in \mathbb{C}^{{N}\times 1}$ are channels between user $1$ and user $2$, RIS and user $2$, and user $1$ and RIS, respectively. $n_2$ is the white Gaussian noise following $n_2\thicksim \mathcal{C}\mathcal{N}\left( 0,\sigma _{2}^{2} \right) $.
As such, the SINR of decoding the common stream $s_0$ at user $2$ is equal to 
\begin{equation}
\gamma_{c,2}=\frac{\left|  \left( \mathbf{g}_{2}^{H}+\mathbf{h}_{2}^{H}\boldsymbol{\Theta} \mathbf{G} \right) \mathbf{w}_0+\sqrt{P_{r}}\left( h^{}_{1,2}+\mathbf{h}_{r,2}^{H}\boldsymbol{\Theta}\mathbf{h}_{1,r} \right) \right|^2}{\sum\nolimits_{i=1}^2{\left| \left( \mathbf{g}_{2}^{H}+\mathbf{h}_{2}^{H}\boldsymbol{\Theta}\mathbf{G} \right) \mathbf{w}_i \right|^2}+\sigma_{2}^{2}}.
\end{equation}
Likewise, we can arrive at the rate of decoding the common stream $s_0$ as 
$
 r_{c,2}=
\log _2\left( 1+\gamma_{c,2}\right)$.
According to the principle of SIC,  the rate of decoding the private stream $s_2$ can be expressed as 
\begin{equation}
r_{p,2}=\log _2\left( 1+\frac{\left| \left( \mathbf{g}_{2}^{H}+\mathbf{h}_{2}^{H}\mathbf{\Theta G} \right) \mathbf{w}_2 \right|^2}{\left| \left( \mathbf{g}_{2}^{H}+\mathbf{h}_{2}^{H}\mathbf{\Theta G} \right) \mathbf{w}_1 \right|^2+\sigma _{2}^{2}} \right) .
\end{equation}
To make sure that the common message stream $s_0$ is successfully decoded by user 1 and user 2, the achievable rate of $s_0$ shall be subject to the constraint of 
\begin{equation}
r_{\text{C,F}}=\min \left( r_{c,1},r_{c,2} \right).
\end{equation}

\section{Problem Formulation}
Based on the preceding analysis, we can calculate the total information transmission rate for the proposed C-RSMA transmission
framework with the assistance of single-connected RIS. For our proposed framework, user 1 serves as an FD relay, and the sum rate can be expressed as 
\begin{equation}
R_{\text{tot,FD}}=r_{\text{C,F}}+r_{p,1}+r_{p,2}.
\end{equation}
The goal of this work is to maximize the sum-rate of above proposed framework by jointly optimizing the reflective beamforming $\mathbf{\Theta }$ of RIS and the transmit beamforming $\mathbf{W}$ of C-RSMA transmission network. The sum-rate maximizing problem for the downlink RIS-enabled C-RSMA transmission network can be formulated as 
\begin{subequations}
\small
\label{eq:023}
\begin{align}
\left( \mathcal{P}_1 \right) \quad &\underset{\mathbf{W},\boldsymbol{\Theta}}{\max}\quad R_{\text{tot,FD}}\\
\text{s.t.}~&\mathrm{Tr}\Big( \mathbf{WW}^H \Big) \le P_{t},\\
&\boldsymbol{\Theta}=\text{diag}( e^{j\theta _1},\ldots ,e^{j\theta _{N}} ), 
~\theta _n\in \left[ 0,2\pi \right) ,\forall n\in \mathcal{N}.
\end{align}
\end{subequations}

Note that the problem $\left( \mathcal{P}_1 \right) $ is a challenging NP-hard problem as both the optimization variables, i.e., the transmit beamformer $\mathbf{W}$ and the reflective beamformer $\boldsymbol{\Theta}$ are coupled together for each time slot. In the following, we propose a low-complexity alternative optimization framework for approximating the optimal solution to this non-convex problem $\left( \mathcal{P}_1 \right) $.

\section{Proposed Joint Transmit and Reflective Beamforming Optimization Framework}
\label{Seciv:Pro}
%
%
\subsection{Optimizing Transmit Beamforming of C-RSMA}
Consider the scenario that user 1 acts as an FD relay. For notational simplicity, we first denote the cascade channel between user $k$ and BS as $\bar{\mathbf{g}}_k = \mathbf{g}_{k}+\mathbf{G}^{H}\boldsymbol{\Theta}^{H} \mathbf{h}_{k}$ and define $\dot{\mathbf{W}}_{k} = \mathbf{w}_{k}\mathbf{w}_{k}^{H}$. Then, the rates of decoding the private stream associated with users 1 and 2 can be re-written by
$r_{p,1}
= \log _2( 1+\dot{\gamma}_{p,1} )$, and
$r_{p,2} 
= \log _2(1+\dot{\gamma}_{p,2} )$, respectively. Herein, $\dot{\gamma}_{p,1}$ and $\dot{\gamma}_{p,2}$ can be expressed as 
\begin{equation}
\small
\dot{\gamma}_{p,1} = \frac{ \bar{\mathbf{g}}^{H}_{1} \dot{\mathbf{W}}_{1} \bar{\mathbf{g}}_{1}}{ \bar{\mathbf{g}}^{H}_{1} \dot{\mathbf{W}}_2 \bar{\mathbf{g}}_{1} +{P}_{r}\left| h_{_{\text{S,I}}}^{} \right|^2+\sigma _{1}^{2}}, \;\,   \dot{\gamma}_{p,2} = \frac{\bar{\mathbf{g}}^{H}_{2} \dot{\mathbf{W}}_2 \bar{\mathbf{g}}_{2} }{ \bar{\mathbf{g}}_{2}^{H} \dot{\mathbf{W}}_1 \bar{\mathbf{g}}_{2} +\sigma _{2}^{2}} .
\end{equation}
Subsequently, we introduce a real-valued auxiliary variable $t$, and with given reflective beamformer $\boldsymbol{\Theta}$, solving the problem   
$\left( \mathcal{P}_1 \right)$ with respect to (w.r.t.) $\mathbf{W}$ amounts to addressing the problem of 
\begin{subequations}
\small
\label{eq:025}
\begin{align}
\left( \mathcal{P}_{1.1} \right) \quad &\underset{\mathbf{W},~t}{\max}\quad t+\sum_{k=1}^{2}r_{p,k}\\
\text{s.t.}~& 
r_{c,k} \geq t, \; \forall k\in\{1,2\},~\mathrm{Tr}\left(\sum\nolimits_{l=0}^{2}\dot{\mathbf{W}}_{l} \right) \le P_{t}.
\end{align}
\end{subequations}
Due to the multiple fractions involved in the objective function in (\ref{eq:025}a) and the non-convexity of the constraint (\ref{eq:025}b), problem $\left( \mathcal{P}_{1.1} \right)$ is highly
intractable and non-convex. To conquer this challenge, we resort to the semidefinite relaxation (SDR) and successive convex approximation (SCA) techniques. To be concrete, we first rewrite the object function $\sum\nolimits_{k=1}^2r_{p,k}$ as
\begin{align}
\small
\begin{split}
\sum\limits_{k=1}^2 r_{p,k} =& \log _2\Big(
\sum\limits_{i=1}^{2} \bar{\mathbf{g}}^{H}_{1} \dot{\mathbf{W}}_i
\bar{\mathbf{g}}^{}_{1} +{P}_{{r}} \left| {h}_{\text{S,I}}^{} \right|^2+\sigma _{1}^{2}\Big)  + \\ & \log _2\Big(
\sum\limits_{j=1}^{2} \bar{\mathbf{g}}^{H}_{2} \dot{\mathbf{W}}_j
\bar{\mathbf{g}}^{}_{2} + \sigma _{2}^{2}\Big) -
\log _2\Big(
\bar{\mathbf{g}}^{H}_{1} \dot{\mathbf{W}}_2
\bar{\mathbf{g}}^{}_{1}  \\ & + {P}_{{r}}\left| {h}_{\text{S,I}}^{} \right|^2+\sigma _{1}^{2}\Big) -  \log _2\Big(
\bar{\mathbf{g}}^{H}_{2} \dot{\mathbf{W}}_1
\bar{\mathbf{g}}^{}_{2} + \sigma _{2}^{2}\Big).
\end{split}
\end{align}
Then, we observe that both logarithmic functions are concave w.r.t. $\mathbf{W}$. Thus, we approximate the above objective functions $ r_{p,1}$ and $ r_{p,2}$ exploiting the iterative SCA technique, respectively. By using the first-order Taylor expansion, we have 
\begin{equation}
\small
\label{eq:029}
\begin{split}
 & \log _2\Big(
\bar{\mathbf{g}}^{H}_{1} \dot{\mathbf{W}}_2
\bar{\mathbf{g}}^{}_{1}  + {P}_{r}\left| h_{\text{S,I}}^{} \right|^2+\sigma _{1}^{2}\Big) \\
 & \leq a_{l}^{(n-1)}+\frac{\log_{2}\mathrm{e}}{2^{a_{l}^{(n-1)}}}\bar{\mathbf{g}}^{H}_{1}\left(\dot{\mathbf{W}}_{2}-\dot{\mathbf{W}}_{2}^{(n-1)}\right)\bar{\mathbf{g}}_{1} \\
 & \triangleq\underline{r}_{p,1}, 
 \end{split}
\end{equation}
where $\dot{\mathbf{W}}_{2}^{(n-1)}$ is the solution to $\dot{\mathbf{W}}_{2}$ in the $(n-1)$-th iteration, and $a_{l}^{(n-1)} = \log _2( \bar{\mathbf{g}}^{H}_{1} \dot{\mathbf{W}}_{2}^{(n-1)}
\bar{\mathbf{g}}^{}_{1}  + {P}_{r}| h_{\text{S,I}} |^2+\sigma _{1}^{2} )$. Hence, we can arrive at that 
\begin{equation}
\small
    r_{p,1} \geq  \log _2\Big(
\sum\nolimits_{i=1}^{2} \bar{\mathbf{g}}^{H}_{1} \dot{\mathbf{W}}_i
\bar{\mathbf{g}}^{}_{1} + {P}_{r}| h_{\text{S,I}}^{} |^2+\sigma _{1}^{2}\Big) - \underline{r}_{p,1},
\end{equation}
which in practice is a lower bound of $ r_{p,1} $ and is concave w.r.t. $\mathbf{W}$.
Similar to (\ref{eq:029}), we can obtain the optimal solution to $\dot{\mathbf{W}}_{1}$ by exploiting the iterative SCA technique against the objective function $r_{p,2}$, i.e., 
\begin{equation}
\small
\label{eq:0031}
\begin{split}
 & \log _2\Big(
\bar{\mathbf{g}}^{H}_{2} \dot{\mathbf{W}}_1
\bar{\mathbf{g}}^{}_{2} + \sigma _{2}^{2}\Big) \\
 & \leq b_{l}^{(n-1)}+\frac{\log_{2}\mathrm{e}}{2^{b_{l}^{(n-1)}}}\bar{\mathbf{g}}^{H}_{2}\left(\dot{\mathbf{W}}_{1}-\dot{\mathbf{W}}_{1}^{(n-1)}\right)\bar{\mathbf{g}}_{2} \\
 & \triangleq\underline{r}_{p,2}, 
 \end{split}
\end{equation}
where $\dot{\mathbf{W}}_{1}^{(n-1)}$ is the solution to $\dot{\mathbf{W}}_{1}$ in the $(n-1)$-th iteration, and $b_{l}^{(n-1)} = \log_2( \bar{\mathbf{g}}^{H}_{2} \dot{\mathbf{W}}_{1}^{(n-1)}
\bar{\mathbf{g}}^{}_{2}  +\sigma _{2}^{2} )$. Herein, $\underline{r}_{p,2}$ is the lower bound of the rate for decoding the private stream associated with the user $2$, thereby is concave w.r.t. $\mathbf{W}$.

Next, we handle the non-convex constraint (\ref{eq:025}b). Specifically, we rewrite the constraint (\ref{eq:025}b) as 
\begin{equation}
\small
\label{eq:031}
    \frac{\bar{\mathbf{g}}_{1}^{H} \dot{\mathbf{W}}_0  \bar{\mathbf{g}}_{1}} {\sum\nolimits_{i=1}^2{\bar{\mathbf{g}}_{1}^{H} \dot{\mathbf{W}}_i  \bar{\mathbf{g}}_{1} + P_{r}|h_{\text{S,I}}^{} |^2 +\sigma_{1}^{2}}}  \geq  2^{t}-1,
\end{equation}
\begin{equation}
\small
\label{eq:032}
\frac{\left| \bar{\mathbf{g}}^{H}_2 \mathbf{w}_0   + \sqrt{P_{r}}\left( h_{1,2}^{}+\mathbf{h}_{r,2}^{H}\boldsymbol{\Theta h}_{1,r} \right)  \right|^2}{\sum\nolimits_{i=1}^2{\bar{\mathbf{g}}^{H}_2 \dot{\mathbf{W}}_i \bar{\mathbf{g}}_2}+ \sigma_{2}^{2}}  \geq 2^t-1.
\end{equation}
Then, we use the arithmetic geometric mean (AGM) inequality to approximate the non-convex constraint (\ref{eq:031}) \cite{9508885},\cite{10757612}. Thus, we have 
\begin{equation}
\small
\label{eq:033}
\begin{split}
        (d^{(n)}\cdot (2^{t}-1))^2+\left(\frac{\sum\nolimits_{i=1}^2{\bar{\mathbf{g}}_{1}^{H} \dot{\mathbf{W}}_i  \bar{\mathbf{g}}_{1} + P_{r}\left|h_{\text{S,I}}^{} \right|^2 }}{d^{(n)}}\right)^2\\
        \leq2\bar{\mathbf{g}}_{1}^{H} \dot{\mathbf{W}}_0  \bar{\mathbf{g}}_{1}-(2^{t+1}-2)\sigma_1^2,
\end{split}
\end{equation}
where 
$
d^{(n)}=\sqrt{\frac{(\sum_{i=1}^2{\bar{\mathbf{g}}_{1}^{H} \dot{\mathbf{W}}_i  \bar{\mathbf{g}}_{1} + P_{r} |h_{_{\text{S,I}}} |^2})^{(n-1)}}{(2^{t}-1)^{(n-1)}}}
$
is the value of $d$ in the $n$-th iteration \cite{10757612}. Similar to~(\ref{eq:033}), we can handle the non-convex constraint (\ref{eq:032}). For notational simplicity, we first let $\bar{h}_{1,2} = h_{1,2}^{}+\mathbf{h}_{r,2}^{H}\boldsymbol{\Theta h}_{1,r}$. Then, the term $\left| \bar{\mathbf{g}}^{H}_2 \mathbf{w}_0   + \sqrt{P_{r}} \left( h_{1,2} + \mathbf{h}_{r,2}^{H}\boldsymbol{\Theta h}_{1,r} \right)  \right|^2$ in (\ref{eq:032}) can be reduced to $\bar{\mathbf{w}}_{0}^{H} \boldsymbol{\Omega}\bar{\mathbf{w}}_{0} + |\sqrt{P_{r}}\bar{h}_{1,2}|^2$, where $\bar{\mathbf{w}}_{0} = [\mathbf{w}_{0};1]$ and 
\begin{equation}
\small
    \label{eq:0320026}
    \boldsymbol{\Omega}  =
    \left.\left[
\begin{array}
{cc}{\mathbf{\bar{g}_{2}}\mathbf{\bar{g}_{2}}^{H}
} & {\big(\sqrt{P_{r}}\bar{h}_{1,2}\big)^{*}\mathbf{\bar{g}_{2}}} \\
{{\big(\sqrt{P_{r}}\bar{h}_{1,2}\big)^{}\mathbf{\bar{g}}_{2}^{H}}}  & {0}
\end{array}\right.\right] .
\end{equation}
Likewise, we use the AGM inequality to approximate the non-convex constraint (\ref{eq:032}). Thus, we have 
\begin{equation}
\small
\label{eq:00033}
\begin{split}
        (\dot{d}^{(n)}\cdot (2^{t}-1))^2+\left(\frac{\sum\nolimits_{i=1}^2{\bar{\mathbf{g}}_{2}^{H} \dot{\mathbf{W}}_i  \bar{\mathbf{g}}_{2} }}{\dot{d}^{(n)}}\right)^2\\
        \leq2(\mathrm{Tr}( \boldsymbol{\Omega} \bar{\mathbf{w}}_{0} \bar{\mathbf{w}}_{0}^{H}) + |\sqrt{P_{r}}\bar{h}_{1,2}|^2 )-(2^{t+1}-2)\sigma_2^2,
\end{split}
\end{equation}
where 
$
\dot{d}^{(n)}=\sqrt{\frac{(\sum\nolimits_{i=1}^2{\bar{\mathbf{g}}_{2}^{H} \dot{\mathbf{W}}_i  \bar{\mathbf{g}}_{2} })^{(n-1)}}{(2^{t}-1)^{(n-1)}}}
$ is the value of $\dot{d}$ in the $n$-th iteration \cite{10757612}. 

In this case, tackling problem $\left(\mathcal{P}_{1.1}\right)$ amounts to solving the relaxed problem $\left(\mathcal{P}_{1.2}\right)$ during the $n$-th update, i.e.,  
\begin{subequations}
\small
\begin{align*}
\label{eq:028}
\left( \mathcal{P}_{1.2} \right) \quad &\underset{\mathbf{W},~t}{\max}\quad t+\log _2\Big(
\sum\nolimits_{i=1}^{2} \bar{\mathbf{g}}^{H}_{1} \dot{\mathbf{W}}_i
\bar{\mathbf{g}}^{}_{1} + {P}_{{r}} \left| h_{\text{S,I}}^{} \right|^2+\sigma _{1}^{2}\Big)\\ & \quad\quad\quad\quad\quad + \log _2\Big(
\sum\nolimits_{j=1}^{2} \bar{\mathbf{g}}^{H}_{2} \dot{\mathbf{W}}_j
\bar{\mathbf{g}}^{}_{2} + \sigma _{2}^{2}\Big) - \underline{r}_{p,1} - \underline{r}_{p,2}\\
\text{s.t.}~
& (\ref{eq:025}b), (\ref{eq:033}), (\ref{eq:00033}).
\end{align*}
\end{subequations}
It is evident that the problem $\left(\mathcal{P}_{1.2} \right)$ is a convex problem and can be handled by the existing convex optimization
algorithms or toolboxes, e.g., CVX \cite{4012317}. 

\subsection{Optimizing Reflective Beamforming of RIS}
With given transmit beamforming vectors $\mathbf{w}_i$ of C-RSMA, $\forall i\in\{0,1,2\}$, the original signal-connected RIS-enabled C-RSMA transmission network sum-rate optimization $\left( \mathcal{P}_1 \right)$ w.r.t. the reflective beamforming $\boldsymbol{\Theta}$ then can be reformulated as 
\begin{subequations}
\small
\label{eq:026}
\begin{align}
\left( \mathcal{P}_{1.3} \right) \quad &\underset{\boldsymbol{\Theta},~t}{\max}\quad t+r_{\text{p,}1}+r_{\text{p,}2}\\
\text{s.t.}~& 
r_{c,k} \geq t, \; \forall k\in\{1,2\}, \\
&(\ref{eq:023}c).
\end{align}
\end{subequations}
It is evident that the above optimization problem is intractable and NP-hard due to the non-convexity of the objective function (\ref{eq:026}a) and the inequality constraint (\ref{eq:026}b). 

To conquer this challenge, we leverage the SDR and the SCA techniques. By introducing $v_n = e^{j\theta_{n}}$, $\forall n\in\mathcal{N}$, $\mathbf{v} = [v_1,v_2,\ldots,v_{N}]^{T}$, we obtain that $\mathbf{h}_{k}^{H}\boldsymbol{\Theta}\mathbf{G}\mathbf{w}_{i} = \mathbf{m}_{k,i}^{H}\mathbf{v}$, where 
$\mathbf{w}_{i}^{H}\mathbf{G}^{H}\otimes \mathbf{h}_{k}^{H} = \mathbf{m}_{k,i}^{H} \in \mathbb{C}^{1\times N}$. By defining $h_{k,k} = \mathbf{g}_{k}^{H} \mathbf{w}_{k}$, $\left| \left( \mathbf{g}_{1}^{H}+\mathbf{h}_{1}^{H}\boldsymbol{\Theta }\mathbf{G} \right) \mathbf{w}_1 \right|^2$ then can be re-expressed as $\left|\mathbf{v}_{1}^{H}\mathbf{m}_{k,k} + h_{k,k}\right|^{2}$. To simplify the calculation, we further reduce this term $\left|\mathbf{v}_{1}^{H}\mathbf{m}_{k,k} + h_{k,k}\right|^{2}$ to $\bar{\mathbf{v}}_{1}^{H}\mathbf{L}_{k,k}\bar{\mathbf{v}}_{1} + \left|h_{k,k}\right|^{2}$, where $\bar{\mathbf{v}}_{1} = [\mathbf{v}_{1}; 1]$ and 
\begin{equation}
\small
\label{eq:0037}
    \left.\mathbf{L}_{k,k}=\left[
\begin{array}
{ll}{\mathbf{m}_{k,k}\mathbf{m}_{k,k}^{H}} & {\mathbf{m}_{k,k}h_{k,k}} \\
{h_{k,k}^{*}\mathbf{m}_{k,k}^{H}} & {0}
\end{array}\right.\right].
\end{equation}
Hence, we have that $\bar{\mathbf{v}}_{1}^{H}\mathbf{L}_{k,k}\bar{\mathbf{v}}_{1} = \mathrm{Tr}(\mathbf{L}_{k,k}\bar{\mathbf{v}}_{1}\bar{\mathbf{v}}_{1}^{H})$. By using the SDR technique, $\sum\nolimits_{k=1}^2 r_{p,k}$ can be re-written by 
$
\log _2( 1+\bar{\gamma}_{p,1} ) + \log _2( 1+\bar{\gamma}_{p,2} )
$, in which $\bar{\gamma}_{p,1}$ and $\bar{\gamma}_{p,2}$ are given by
\begin{equation}
\small
    \bar{\gamma}_{p,1} = \frac{\mathrm{Tr}(\mathbf{L}_{1,1}\bar{\mathbf{v}}_{1}\bar{\mathbf{v}}_{1}^{H}) + \left|h_{1,1}\right|^2}{
\mathrm{Tr}(\mathbf{L}_{1,2}\bar{\mathbf{v}}_{1}\bar{\mathbf{v}}_{1}^{H}) + \left|h_{1,2}\right|^2
+ P_{r}\left| h_{_{\text{S,I}}} \right|^2+\sigma _{1}^{2}}, 
\end{equation}
\begin{equation}
    \small
    \bar{\gamma}_{p,2} = \frac{\mathrm{Tr}(\mathbf{L}_{2,2}\bar{\mathbf{v}}_{1}\bar{\mathbf{v}}_{1}^{H}) + \left|h_{2,2}\right|^2
}{\mathrm{Tr}(\mathbf{L}_{2,1}\bar{\mathbf{v}}_{1}\bar{\mathbf{v}}_{1}^{H}) + \left|h_{2,1}\right|^2 +\sigma_{2}^{2}}.
\end{equation}
Since the above objective function is still complicated and non-convex, we further leverage the SCA technique to relax it. Specifically,  $\sum\nolimits_{k=1}^2 r_{p,k}$ can be re-written by
\begin{equation}
\label{eq:039}
\small
\begin{split}
\sum\limits_{k=1}^2 r_{p,k} =& 
\log _2\Big( \sum_{i=1}^{2}
\big(
\text{Tr}\big(\mathbf{L}_{1,i}\bar{\mathbf{v}}_{1}\bar{\mathbf{v}}_{1}^{H}\big)
 + \left|h_{1,i}\right|^2\big) +  P_{r}\left| h_{\text{S,I}} \right|^2+\sigma _{1}^{2} \Big) 
\\& + 
 \log _2\left( 
\sum_{j=1}^{2} \big(
\text{Tr}\big(\mathbf{L}_{2,j}\bar{\mathbf{v}}_{1}\bar{\mathbf{v}}_{1}^{H}\big) + \left|h_{2,j}\right|^2\big) +\sigma_{2}^{2}\right)
- \\&
\log_2\Big(
\text{Tr}\big(\mathbf{L}_{1,2}\bar{\mathbf{v}}_{1}\bar{\mathbf{v}}_{1}^{H}\big) + \left|h_{1,2}\right|^2 +
P_{r}\left| h_{\text{S,I}} \right|^2+\sigma _{1}^{2}
\Big)
\\& -\log _2\Big( 
\text{Tr}\big(\mathbf{L}_{2,1}\bar{\mathbf{v}}_{1}\bar{\mathbf{v}}_{1}^{H}\big) + \left|h_{2,1}\right|^2 +\sigma_{2}^{2} \Big).
\end{split}
\end{equation}

Next, we handle the non-convex constraint (\ref{eq:026}b) by using the AGM inequality
to approximate the constraint (\ref{eq:026}b). Specifically, we have
\begin{equation}
\small
\label{eq:040}
    \frac{\mathrm{Tr}(\mathbf{L}_{1,0}\bar{\mathbf{v}}_{1}\bar{\mathbf{v}}_{1}^{H}) + \left|h_{1,0}\right|^2}{
\sum\nolimits_{i=1}^{2}(\mathrm{Tr}(\mathbf{L}_{1,i}\bar{\mathbf{v}}_{1}\bar{\mathbf{v}}_{1}^{H}) + \left|h_{1,i}\right|^2)
+ P_{r}\left| h_{\text{S,I}} \right|^2+\sigma _{1}^{2}}   \geq  2^{t}-1,
\end{equation}
\begin{equation}
\small
\label{eq:041}
\frac{\left|  ( \mathbf{g}_{2}^{H}+\mathbf{h}_{2}^{H}\mathbf{\Theta G} ) \mathbf{w}_0+ \sqrt{P_{r}}( h_{1,2}+\mathbf{h}_{r,2}^{H}\boldsymbol{\Theta }\mathbf{h}_{1,r} )   \right|^2
}{\sum\nolimits_{j=1}^{2}(\mathrm{Tr}(\mathbf{L}_{2,j}\bar{\mathbf{v}}_{1}\bar{\mathbf{v}}_{1}^{H}) + \left|h_{2,j}\right|^2) +\sigma_{2}^{2}}  \geq 2^t-1.
\end{equation}
Afterwards, we use the AGM inequality to approximate the non-convex constraint (\ref{eq:026}b). Thus, we get  
\begin{equation}
\small
\label{eq:042}
\begin{split}
        (z^{(n)} (2^{t}-1))^2&+\left(\frac{\sum\nolimits_{i=1}^{2}\big(\mathrm{Tr}(\mathbf{L}_{1,i}\bar{\mathbf{v}}_{1}\bar{\mathbf{v}}_{1}^{H}) + \left|h_{1,i}\right|^2\big)
+ P_{r}\left| h_{\text{S,I}} \right|^2}{z^{(n)}}\right)^2\\&
        \leq2\mathrm{Tr}\Big(\mathbf{L}_{1,0}\bar{\mathbf{v}}_{1}\bar{\mathbf{v}}_{1}^{H}\Big) + 2\left|h_{1,0}\right|^2-(2^{t+1}-2)\sigma_1^2,
\end{split}
\end{equation}
where 
$
{z}^{(n)}=\sqrt{\frac{\big(\sum_{i=1}^{2}(\mathrm{Tr}(\mathbf{L}_{1,i}\bar{\mathbf{v}}_{1}\bar{\mathbf{v}}_{1}^{H}) + \left|h_{1,i}\right|^2)
+ P_{r}\left| h_{_{\text{S,I}}} \right|^2\big)^{(n-1)}}{(2^{t}-1)^{(n-1)}}}
$ is the value of $z$ in the $n$-th iteration \cite{10757612}. 
Afterwards, we handle the non-convex constraint (\ref{eq:041}). Specifically, we transfer the linear items $\left(\mathbf{g}_{2}^{H}+\mathbf{h}_{2}^{H}\mathbf{\Theta G} \right) \mathbf{w}_0$ and $\sqrt{P_r}\left(h_{1,2}+\mathbf{h}_{r,2}^{H}\boldsymbol{\Theta }\mathbf{h}_{1,r}\right)$ in the numerator of formula (\ref{eq:041}) into the matrix form. In addition, we define $\mathbf{v}_2 = \text{vec}(\boldsymbol{\Theta})$, 
$\bar{\mathbf{v}}_{2} = [\mathbf{v}_2; 1]$. In this case, we reduce the above terms $\mathbf{h}_{2}^{H}\mathbf{\Theta G}  \mathbf{w}_0$ and $\mathbf{h}_{r,2}^{H}\boldsymbol{\Theta }\mathbf{h}_{1,r}$ to the linear matrix forms $\mathbf{w}_0^{H} \mathbf{G}^{H} \otimes \mathbf{h}_{2}^{H}\mathbf{v}_{2}$ and $\mathbf{h}_{1,r}^{T}\otimes \mathbf{h}_{r,2}^{H}\mathbf{v}_{2}$, respectively. Hence, we have 
\begin{equation}
\small
\label{eq:00043}
\begin{split}
&\left|  \left( \mathbf{g}_{2}^{H}+\mathbf{h}_{2}^{H}\mathbf{\Theta G} \right) \mathbf{w}_0+ \sqrt{P_{r}}\left( h_{1,2}+\mathbf{h}_{r,2}^{H}\boldsymbol{\Theta }\mathbf{h}_{1,r} \right) \right|^2 \\&=
 \bar{\mathbf{v}}_{2}^{H} 
    \boldsymbol{\Gamma}_{k,k}
    \bar{\mathbf{v}}_{2} + |\mathbf{g}_{2}^{H}\mathbf{w}_{0}+\sqrt{P_{r}}h_{1,2}|^{2}\\ &=
    \mathrm{Tr}\Big(\boldsymbol{\Gamma}_{k,k} \bar{\mathbf{v}}_{2} \bar{\mathbf{v}}_{2}^{H}\Big) + |\mathbf{g}_{2}^{H}\mathbf{w}_{0}+\sqrt{P_{r}}h_{1,2}|^{2},
\end{split}
\end{equation}
where $\mathbf{w}_0^{H} \mathbf{G}^{H} \otimes \mathbf{h}_{2}^{H}+ \sqrt{P_{r}}(\mathbf{h}_{1,r}^{T}\otimes \mathbf{h}_{r,2}^{H}) = \boldsymbol{\varpi}^{H}$, and  

\begin{equation}
\label{eq:00044}
\small
    \boldsymbol{\Gamma}_{k,k} =
    \left.\left[
\begin{array}
{cc}{\boldsymbol{\varpi}\boldsymbol{\varpi}^{H}
} & {\big(\mathbf{g}_{2}^{H}\mathbf{w}_{0}+\sqrt{P_{r}}h_{1,2}\big)^{*}\boldsymbol{\varpi}} \\
{\big(\mathbf{g}_{2}^{H}\mathbf{w}_{0}+\sqrt{P_{r}}h_{1,2}\big)\boldsymbol{\varpi}^{H}}  & {0}
\end{array}\right.\right] .
\end{equation}

Then, substituting (\ref{eq:00043}) and (\ref{eq:00044}) back into (\ref{eq:041}) 
and invoking the AGM inequality to approximate the non-convex constraint (\ref{eq:041}) yields 
\begin{equation}
\small
\label{eq:0044}
\begin{split}
       & (\dot{z}^{(n)} \cdot (2^{t}-1))^2+\left(\frac{\sum\nolimits_{j=1}^{2}\big(\mathrm{Tr}(\mathbf{L}_{2,j}\bar{\mathbf{v}}_{1}\bar{\mathbf{v}}_{1}^{H}) + \left|h_{2,j}\right|^2\big)
}{\dot{z}^{(n)}}\right)^2\\&
        \leq2\mathrm{Tr}\big(\boldsymbol{\Gamma}_{k,k}\bar{\mathbf{v}}_{2}\bar{\mathbf{v}}_{2}^{H}\big) + 2|\mathbf{g}_{2}^{H}\mathbf{w}_{0}+\sqrt{P_{r}}h_{1,2}|^{2}-(2^{t+1}-2)\sigma_2^2,
\end{split}
\end{equation}
where 
$
\dot{z}^{(n)}=\sqrt{\frac{(\sum\nolimits_{j=1}^{2}\mathrm{Tr}(\mathbf{L}_{2,j}\bar{\mathbf{v}}_{1}\bar{\mathbf{v}}_{1}^{H}) + \left|h_{2,j}\right|^2)^{(n-1)}}{(2^{t}-1)^{(n-1)}}}
$ is the value of $\dot{z}$ in the $n$-th iteration \cite{10757612}. 

Given the above approximations and substituting the semidefinite relaxed object function $\sum\nolimits_{k=1}^2 r_{p,k}$ into (\ref{eq:026}), the problem $\left( \mathcal{P}_{1.3} \right) $ can be further recast into the optimization problem during the $n$-th update, i.e., 
\begin{subequations}
\small
\label{eq:0043}
\begin{align}
\left( \mathcal{P}_{1.4} \right) \; &\underset{\bar{\mathbf{v}}_{1}, \bar{\mathbf{v}}_{2},~t}{\max}\quad t+r_{p,1}+r_{p,2}\\
\text{s.t.} \quad & 
 (\ref{eq:023}c), (\ref{eq:042}), (\ref{eq:0044}).
\end{align}
\end{subequations}
Notably, the above relaxed problem $\left( \mathcal{P}_{1.4} \right)$ becomes to be a convex problem and can be effectively tackled by using the existing convex
optimization problem solver, i.e., CVX~\cite{4012317}.

\subsection{Convergence and Complexity Analysis}
The proposed alternative optimization algorithm begins with initial transmit beamforming matrix $\mathbf{W}^{(0)}$, RIS phase matrix $\boldsymbol{\Theta}^{(0)}$, and auxiliary variables $t^{(0)}$, $d^{(0)}$, $\dot{d}^{(0)}$, $z^{(0)}$, $\dot{z}^{(0)}$. In the $n$-th iteration, we update $\{\mathbf{W}^{(n)}, t^{(n)},\underline{r}_{p,1}^{(n)}, \underline{r}_{p,2}^{(n)},d^{(n)},\dot{d}^{(n)}\}$ with given $\boldsymbol{\Theta}^{(n-1)}$. Given $\mathbf{W}^{(n)}$, the set $\{\boldsymbol{\Theta}^{(n)}, t^{(n)},r_{p,1}^{(n)},r_{p,2}^{(n)},z^{(n)},\dot{z}^{(n)}\}$ then is updated based on SDR and SCA. The above process repeats until a minimum achievable rate gain of common stream $s_{0}$ within two adjacent iterations, i.e., $\Delta t = \left| \frac{ t^{(n)} -  t^{(n-1)}}{ t^{(n-1)}}\right|$, is achieved.

It is noted that, the computational cost of the proposed low-complexity
alternative optimization approach is incurred mainly for i) the RIS reflection optimization, and ii) the transmit beamforming optimization. More concretely, the proposed scheme reduces the RIS reflection optimization from $\mathcal{O}({N^3})$ to $\mathcal{O}{(N^2)}$ through vectorization operations. On the other hand, the transmit beamforming optimization adopts hierarchical Taylor expansion, decomposing the original non-convex problem into quadratic programming with linear constraints, thus imposing a computational cost of $\mathcal{O}({M_t^3)}$. Compared with the global method that requires joint optimization of $(N+M_t)$ variables, our method separates the computational complexity into independent sub-problems of  $\mathcal{O}({M_t^3)}$ and $\mathcal{O}({N^2)}$.


\section{Simulation Results}
In this section, we evaluate the performance of the proposed approach when the BS and RIS are deployed at coordinates (0, 0) m and (30, 0) m, respectively. User $1$ and user $2$ are located at coordinates (25, -7) m and (35, -7) m, respectively. In this paper, we assume that the RIS-assisted C-RSMA transmission system operates at mmWave frequency band. Thus, we adopt the Saleh
Valenzuela parametric channel model to construct the involved five individual channels $\mathbf{g}_k$, $\mathbf{G}$, $\mathbf{h}_k$, $h_{\text{S,I}}$, and $h_{1,2}$.
For simplicity, we assume that 
the path loss between the source and destination is $\rho_{\text{s,d}} = d_{\text{s,d}}^2$, where $d_{\text{s,d}}$ is the link distance between the source and destination. Herein, the reference distance is $d_0 =$  1 m.
The noise tolerance is $\sigma_{i}^2 = $ -66 dBm, $i\in\{1,2\}$, and the minimum achievable rate gain of common stream $s_{0}$ within two adjacent iterations is $\Delta t =$  $10^{-3}$. Besides, the wavelength is $\lambda =$  $10^{-2}$m.

\begin{figure}[t!] 
    \centering
    \includegraphics[width=7.5cm]{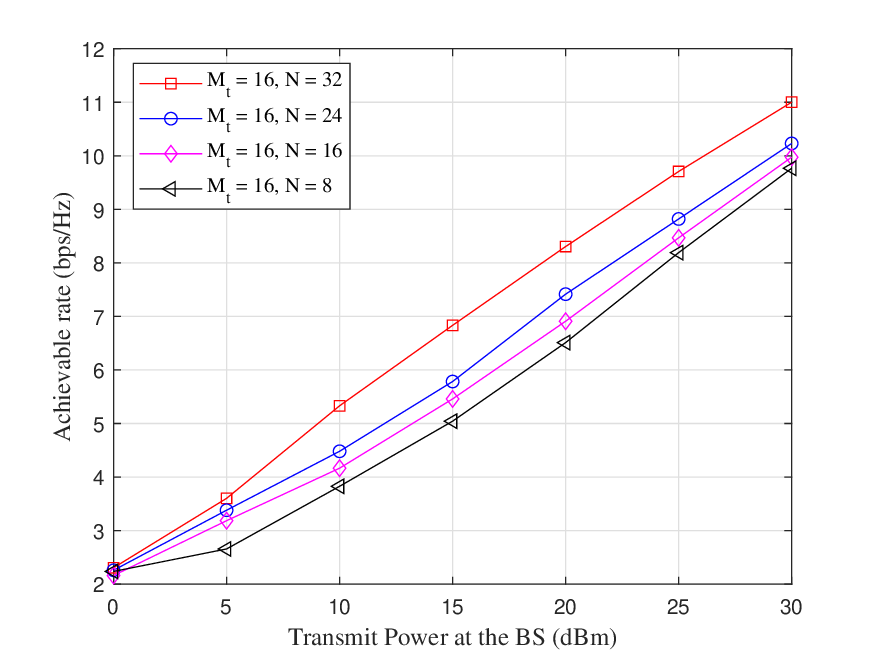}
    \caption{Achievable rate performance of the proposed approach with distinct reflecting element configurations at RIS, and $P_r =$  0 dBm.}
    \label{fig:simu01}
\end{figure}

 For demonstrating the performance benefits of our system, we
 compare our system with the following counterpart schemes:
    i) \textbf{C-RSMA}, which is the proposed cooperative RSMA transmission scheme assisted by RIS and user relaying, as described in Section \ref{Seciv:Pro};
     ii) \textbf{SDMA}, which is a special case of C-RSMA when the power allocated to the common stream is zero;
     iii) \textbf{HD-RSMA}, which integrates the RIS and HD communication into RSMA to sequentially deliver the common stream and private stream to all users \cite{8846761};     
     %
     %
     iv) \textbf{no RIS SDMA}, which is the multi-user linearly precoded SDMA without using RIS.
     %

\begin{figure}[t!] 
    \centering
 \setlength{\belowcaptionskip}{-0.3cm} 
    \includegraphics[width=7.5cm]{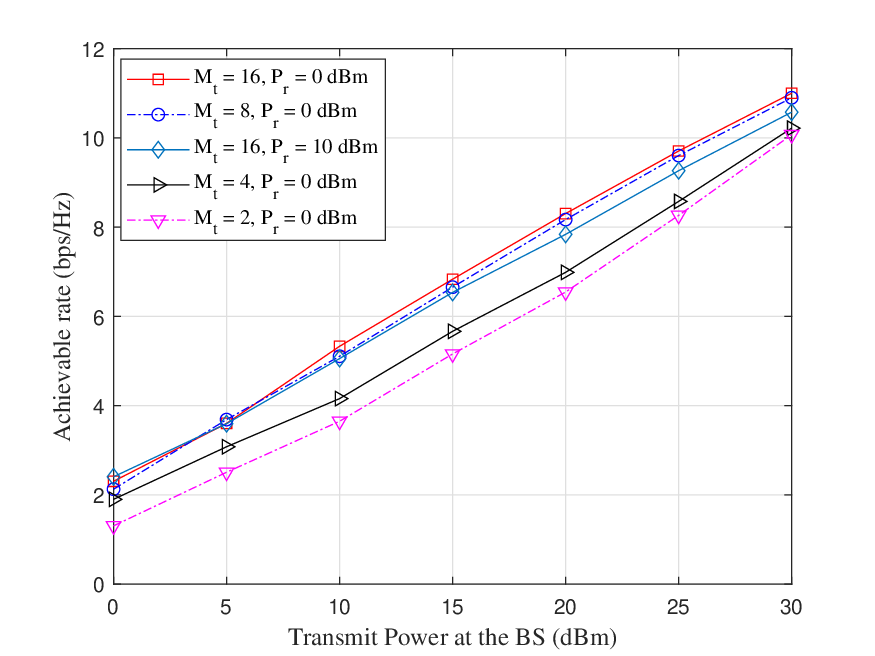}
    \caption{Achievable rate performance of the proposed approach with distinct antenna configurations at BS, $N = 32$, and $P_r =$ 0 or 10 dBm.}
    \label{fig:simu02}
\end{figure}

Fig. \ref{fig:simu01} depicts the achievable rate of the proposed C-RSMA transmission system with $M_t =$ 16 and various configurations of RIS reflecting elements $N$. Obviously, it can be observed that increasing the value of $N$ yields an improved achievable rate performance of our C-RSMA. Meanwhile, this figure indicates that with given achievable rate, the proposed system with larger $N$ consumes lower transmit power at the BS. 
On the other hand, Fig.~\ref{fig:simu02} shows that the achievable rate of C-RSMA with $N =$ 32 and distinct antenna configurations at the BS. We observe from Fig. \ref{fig:simu02} that C-RSMA with $M_t =$ 16 antennas enables higher performance gain than C-RSMA with $M_t =$ 8, 4, and 2 antennas, respectively. Fig. \ref{fig:simu02} also reveals that with given $P_t$, increasing the transmit power for D2D communication leads to performance degradation, particularly at the high-SNR region. 

\begin{figure}[t!] 
    \centering
 \setlength{\belowcaptionskip}{-0.3cm} 
    \includegraphics[width=7.5cm]{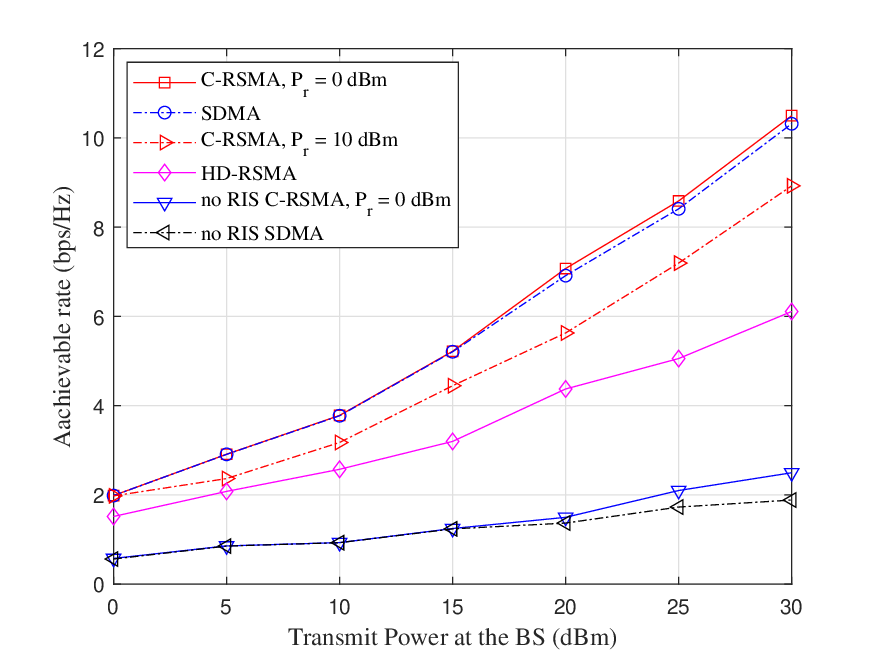}
    \caption{Achievable rate performance comparisons of the proposed C-RSMA, SDMA, HD-RSMA, no RIS C-RSMA, and no RIS SDMA approaches with $M_t =$  $N =$ 4, and $P_r =$ 0 or 10 dBm.}
    \label{fig:simu03}
\end{figure}

Fig. \ref{fig:simu03} portrays the achievable rate performance of the SDMA, HD-RSMA, no RIS C-RSMA, no RIS SDMA, and the proposed C-RSMA systems with $M_t =$  $N =$ 4, and $P_r =$~0 or 10 dBm. Fig. \ref{fig:simu03} shows that C-RSMA provides superior rate performance than the HD-RSMA, no RIS C-RSMA, no RIS SDMA. The main reason is that C-RSMA enables the common rate performance improvement by reflecting the encoded common stream $s_0$ via the single-connected RIS and relaying $s_0$ to the intended user via D2D transmission link at the same time.   
We also observe that, C-RSMA with $P_r =$ 10 enables performance gain up to 31.6 \% and 74.4 \% over HD-RSMA and no RIS C-RSMA 
when $P_t =$ 30 dBm, respectively. 

\section{Conclusion}
In this paper, we presented a novel C-RSMA transmission framework assisted by RIS and user relaying. Specifically, we formulated the common- and private-rate maximization problem 
satisfying the transmit power
constraint.
To conquer this challenge, we proposed a low-complexity algorithm that jointly optimizes the transmit beamformer at the BS and the reflective beamformer at the RIS. 
Simulation results demonstrated that our approach is capable of enabling better sum-rate performance than the selected up-to-date counterpart methods.

\bibliographystyle{IEEEtran}%
\bibliography{bibfile}
\end{document}